\documentclass{aa}
\newcommand{\lesssim} {\; \buildrel < \over \sim \;}
\newcommand{\gtrsim} {\; \buildrel > \over \sim \;}

\begin{document}
   \thesaurus{(02.12.1;  
               08.19.4;   
               13.07.1;   
               13.25.3)}  

\title{Iron line afterglows: general constraints}

\author{G. Ghisellini\inst{1}, D. Lazzati\inst{1,2}\and S. Campana\inst{1} }

\title{Iron line afterglows: general constraints}
 
\institute{Osservatorio Astronomico di Brera, Via Bianchi 46, I--23807
Merate (Lc), Italy
\and
Dipartimento di Fisica, Universit\`a degli Studi di Milano,
Via Celoria 16, I--20133 Milano, Italy}

\date{Received ; accepted }

\maketitle

\begin{abstract}

The discovery of a powerful and transient iron line feature in the 
X--ray afterglow spectra of $\gamma$--ray bursts would be a major 
breakthrough for understanding the nature of their progenitors,
strongly suggesting the presence of a large, iron rich, mass in the
vicinity of the burst event.
Model--independent limits to the size and the mass of the
the iron line emitting region are derived and discussed.
We also discuss how these results can be used to constrain
the amount of beaming or anisotropy of the burst emission.

\end{abstract}


\section{Introduction}
At this meeting it has been reported the possible detection of an iron line
in the X--ray afterglow spectra of two bursts:
GRB 970508 detected by BeppoSAX (Piro et al. 1999, see also these
proceedings) and GRB 970828 by ASCA (Yoshida et al. these proceedings).
The significance of these features is admittedly not extremely
compelling ($\sim 99\%$ and $\sim 98\%$ for the two events, respectively), 
but the implications that they bear are so important
to justify a study on the radiation mechanisms that would produce it
(Lazzati et al. 1999, see also these proceedings).
We can also put strong constraints on the size and mass
of the line emitting region in a model--independent way.
These limits, which are very robust, point towards the
presence, within a distance of $\sim 10^{16}$ cm from the
burst, of at least  $10^{-3}$ $M_\odot$ of iron.
We also stress that these emission line features, observed
in bursts which also have an optical afterglow, strongly suggest
that the line emitting region is not spherically symmetric, 
but must have some degree of anisotropy.
Finally, this constrains the possible anisotropy of the burst
radiation, since the line emitting region samples different line
of sights than our own.

The cosmological parameters will be set throughout this paper to
$H_0 = 65$~km~s$^{-1}$~Mpc$^{-1}$, $q_0=0.5$ and $\Lambda=0$.

\section{General Constraints}
\label{due}
We here prefer to discuss the implications of detecting
an iron line during the X--ray afterglow of a generic
burst, without referring in particular to the two cases mentioned 
in the introduction.
Let us therefore assume that the flux of the X--ray afterglow
is of the order of 10$^{-13}$ erg cm$^{-2}$ s$^{-1}$,
and assume a redshift of $z=1$.
To be visible during the X--ray afterglow emission, 
the emission line should have a comparable flux\footnote{Here and in the
following we parametrize a quantity $Q$ as $Q=10^xQ_x$ and adopt cgs
units.}, $F_{Fe}=10^{-13}F_{Fe,-13}$ 
erg cm$^{-2}$ s$^{-1}$.
This in itself constrains both the amount of line--emitting matter and
the size of the emitting region.

\subsection{Limit to the size}

Assume for simplicity that the emitting region is a homogeneous
spherical shell at a distance $R$ from the $\gamma-$ray burst, with a width
$\Delta R\le R$.
The fluence of the emission line cannot exceed the absorbed
ionizing fluence $q{\cal F}$ (where ${\cal F}$ is the total GRB 
fluence and $q$ is the fraction of it which is absorbed and reprocessed 
into the line).
The observed duration of the emission line cannot be shorter
than the light crossing time of the region $R/c$.
 From this we obtain the limit
\begin{equation}
R\, <3\, \times 10^{18}\,{q\,{\cal F}_{-5}\over F_{Fe,-13}}\quad {\rm cm}
\end{equation}
Since $q$ is at most  $\sim 0.03$, the emitting region is very compact,
ruling out emission from interstellar matter, even assuming 
the large densities appropriate for star forming regions.
We would like to stress that the above limit
{\it is independent of the variability of the line flux}.

\subsection{Limit to the mass}

The total line photons produced at 6.4--6.9 keV in $10^5\, t_5$
seconds, for a GRB located at $z=1$, are 
$\sim 3\times10^{57}\,F_{Fe,-13}\, t_5$. 
Assuming that each iron atom produces $k$ line photons,
this corresponds to $\sim 150\,F_{Fe,-13}\,t_5/k \, M_\odot$ of iron.
The parameter $k$ depends on the details of the assumed scenario, but
we can set some general limits.
Assume in fact that each iron atom can emit photons only when illuminated 
by an ionizing flux, which is provided by
the burst itself or by the high energy tail of the afterglow emission.
Since the burst radiation has enough power to photoionize all the matter
in the vicinity of the progenitor (see e.g. \cite{boe99}),
line photons will be emitted only through the recombination process.
But even if we assume that the recombination is {\it instantaneous},
the value of the parameter $k$ will not be larger than 
the total number of photoionizations an ion can undergo
during the burst and/or the afterglow.
For iron $K$--shell electrons, with cross section $\sigma_{K}=1.2\times
10^{-20}$~cm$^2$ we have:
\begin{equation}
k\, \lesssim \, {{q\, E} \over {4\,\pi\, \epsilon_{ion} \, R^2}} 
\, \sigma_{K} = 
6.5\times10^6 \, {q\, E_{52} \over R^2_{16}}\, 
\left({9.1 \, {\rm keV} \over \epsilon_{ion}}\right)
\end{equation}
\noindent 
where $E$ is the total energy emitted by the burst and/or afterglow
and $\epsilon_{ion}$ the energy of a single ionizing photon.
This upper limit on $k$ translates in a lower limit on the iron mass:
\begin{equation}
M_{\rm Fe} \, \gtrsim \, 2.3\times10^{-5}\, F_{Fe,-13} \, t_5\, 
{R_{16}^2 \over q\, E_{52}}\,\, M_\odot
\end{equation}
which, for a solar iron abundance, yields a total mass
$M \gtrsim 0.013 \,F_{Fe,-13} \, t_5\, R_{16}^2/(q\, E_{52}) \, 
M_\odot$, i.e. a third of a solar mass for $q \sim 0.03$.

Even if these numbers apparently do not rule out reverberation from a 
molecular cloud (\cite{ghi99, mes98}), we remark that the recombination 
time has been assumed negligible in the above discussion. 
At densities typical of molecular clouds the recombination 
time is larger than the burst duration and the value of $k$ cannot exceed
12, set by the photoelectric yield of the iron atom (e.g. \cite{boe99}).

\subsection{Limit on geometry and isotropy}
Optical afterglow emission has been observed in about
half the $\gamma$--ray burst events for which
the X--ray afterglow has been detected.
In particular, the optical afterglow of GRB 970508 lasted
for hundreds of days (Galama et al. 1998).
If the iron line emitting region were spherically symmetric,
it would inevitably stop the fireball and the usual relatively
slow transformation of bulk kinetic energy into radiation
could not take place.
For this reason it is necessary to assume some special geometry
of the iron line emitting material, which has not to interfere
with the observed optical afterglow.
In other words, this region cannot be located along our line of sight, 
but, on the other hand, it has to be illuminated by the burst
emission, in order to produce the iron line feature (e.g. a torus surrounding 
the central region, or a bicone).
Therefore we conclude that the iron line feature is a powerful
tool to know how isotropic the burst emission is.

\section{Discussion}

We have derived some model--independent limits on the 
size and mass of the material responsible for iron line
features, assuming that they can be detected during the 
X--ray afterglow. 
This work has been stimulated by the recent claim of 
detection of such lines, but the validity of our conclusions
are quite general.
The most important implication of these lines, if confirmed,
regard the progenitor of the burst.
In fact it is inescapable to assume that about a solar mass
of (probably iron rich) material is located in the close
vicinity of the event.
Among the proposed models, the Supranova of Vietri and Stella (1998)
is the only one which naturally accounts for this.
This in turn implies that at least a class of $\gamma$--ray burst
have to be associated with supernovae exploded about a month earlier
(assuming that the remnants have a velocity of about $10^4$ km s$^{-1}$).

The possible association of GRB with supernovae has
been investigated recently in detail by \cite{blo98,kip98} and \cite{wan98},
following the explosion of GRB~980425, likely associated with the type Ic
SN 1998bw. 
Among these works, only \cite{wan98} find evidence for a 
connection while the other two limit to a few percent the bursts
possibly associated with supernovae.
In the Supranova scenario, however, the association of 
supernovae with bursts suffer for a time delay which would smear 
the time correlation between the two phenomena. 

Should the iron lines possibly detected in GRB~970508 and GRB~970828
be real and confirmed by other cases, then we have a strong case
for the connection between supernovae and $\gamma$--ray bursts.
The next generation of experiments and satellites, 
such as XMM, AXAF and ASTRO--E, will provide us with the necessary
information to draw more accurate conclusion on the puzzling
problem of the $\gamma$--ray burst progenitor.

\end{document}